# Efficient trajectory design for distant planetary orbiters [*]

Francesca Scala [†] [‡]
*Politecnico di Milano, Via La Masa 34, 20156 Milano, Italy*

Ioannis Gkolias[§]
*University of Thessaloniki, Thessaloniki, Hellas, 54124*

Camilla Colombo[¶]
*Politecnico di Milano, Via La Masa 34, 20156 Milano, Italy*

## I. Introduction

THE investigation of secular perturbations caused by a third-body attractor was widely studied in the past. In the '60s, Lidov and Kozai described the mathematical model for the orbital evolution of a planet's probes under the effect of a third, expressing the results in the Hamiltonian phase space body[1, 2]. The particle oscillations depend on the orbit's initial eccentricity and inclination, and the effect is more evident for highly inclined orbits, such as Highly Elliptical Orbits (HEOs). In recent years, many studies have arisen on the Lidov-Kozai mechanism for different astrophysical applications. The evolution of orbits around Galilean satellites of Jupiter has been studied in several works [3–6], where the effect of Jupiter's attraction and planet's oblateness $J_2$ effect have been investigated. A double-averaged representation of the dynamical environment has been proposed in [3]. Moreover, different orbit families have been analyzed in [7] for a probe orbiting Jupiter's moons Ganymede and Callisto and Saturn's moon Titan, considering an equatorial reference frame. Another example is the study of orbit design for a probe around Mercury, subject to the attraction of the Sun, as proposed in [8]. A different approach has been presented in [9], where the effect of the Solar radiation pressure has been included in the model on top of the non-spherical harmonics and the third-body perturbations for planning future missions to Mercury. Passing to the Earth environment for satellites in HEO, [10–14] developed a Hamiltonian formulation to include the third-body influence, coupled with the zonal effect of $J_2$. The effect of this coupling mechanism has been discussed in [15], notably considering the inclusion of the $C_{22}$ term in the harmonics.

Starting from the Hamiltonian representation of the dynamics in [13, 14], this work proposes an innovative procedure to design fully-analytical maneuvers for post-mission disposal of HEOs satellites, exploiting the third-body perturbations. The Hamiltonian representation has been selected to include the external perturbing effects and to obtain a phase space representation. Notably, the orbit evolution can be described through the variation of double-averaged orbital elements over the orbital periods of the spacecraft and the perturbing bodies around the central planet, as described in [3, 13, 14]. Starting from [13, 14], this work conveys a two-dimensional Hamiltonian representation under the third-body perturbations and the central planet's oblateness. The effect of solar radiation pressure has been neglected in this analysis. The coupling effect of the gravitational attraction of a third-body (e.g., the Sun or a moon) and the planet's oblateness influences the dynamics of probes flying in high-altitude, and highly-elliptic orbit region [14, 16]. Differently from [14, 17], an equivalent representation in the central planet's equatorial reference frame has been proposed in this work. We derive an innovative reduced one-degree-of-freedom (DOF) model describing the satellite's long-term dynamical evolution in the Hamiltonian phase space. This one-DOF model has been developed for an inclined third body. Nonetheless, we observed that in the case of the Earth-Moon-Sun system, the inclined nature of the Moon causes troubles with the elimination of the node of the third body from the Hamiltonian formulation. On the contrary, a non-inclined third body is not subject to this problem regarding the model's accuracy. An example of the non-inclined third body has already been considered for different applications in [7]. However, the potentiality of the methodology relies on the efficient long-term dynamic representation via the reduced Hamiltonian formulation, with no integration of the dynamical equations. From the reduced Hamiltonian, an innovative fully-analytical approach to design end-of-life maneuvers was designed, targeting the disposal trajectory in the phase space of eccentricity and argument of periapsis. Differently from the semi-analytical description in [18], this paper proposes an analytical model for designing the disposal maneuver with no numerical propagation based on the reduced Hamiltonian representation. This method aims to achieve a natural re-entry by exploiting the long-term effect of the orbital perturbations, enhanced by impulsive maneuvers. After the impulsive maneuver, the new orbit conditions lead to a natural increase of the eccentricity until the atmospheric re-entry is reached. The end-of-life disposal design has been developed for two test case scenarios. First, the disposal of Venus' orbiter has been considered, including the coupling effect of the Sun's gravitational attraction and Venus' oblateness. The relatively small inclination of Venus' equatorial plane over the ecliptic plane results in a scenario where the third body (i.e., Sun) lies on the planet's equatorial plane, similar to [7]. The second scenario considers an HEO satellite orbiting the Earth under the influence of the Earth's oblateness and the combined perturbations of the Moon and Sun, starting from the INTEGRAL disposal in [14]. In this scenario, the fully-analytical method for maneuver design produces discordant results compared to the semi-analytical approach. This behavior is mainly due to the node elimination during the setup of the reduced one-DOF Hamiltonian description for the Earth system.

## II. Dynamical Model Formulation

The dynamic of satellites orbiting a planet is discussed in this section, starting from [19]. First, a double-averaged model is implemented for the secular and long-term analysis. Then, node elimination is applied to drop the dependence on the satellite's right ascension of the ascending node to produce a one-DOF Hamiltonian representation. The model is implemented in the planet's equatorial frame considering the planet's oblateness $J_2$ and the third-body perturbation (e.g., Sun and Moon) up to the fourth order, as in [18].

## A. Orbital perturbations

The orbital dynamic of a massless spacecraft can be represented through the Hamiltonian formulation [20]:

$$\mathcal{H} = -\mathcal{H}_{kep} - \mathcal{R} = -\frac{\mu}{2a} - \mathcal{R}_{zonal} - \mathcal{R}_{3b}, \tag{1}$$

The first term represents the Keplerian contribution, with $\mu$ the planet's gravitational parameter and $a$ the satellite's semi-major axis. The term $\mathcal{R}_{zonal}$ represents the expression of zonal effect, via Legendre polynomials [20]:

$$\mathcal{R}_{zonal} = -\frac{\mu}{r} \sum_{l=2}^{\infty} J_l \left(\frac{R_\alpha}{r}\right)^l P_l(\sin \delta), \tag{2}$$

$J_l$ are the zonal harmonic coefficients, $R_\alpha$ is the planet mean equatorial radius, $r$ is the magnitude of the satellite position vector, $P_l(\sin \delta)$ are the the associated Legendre polynomials of degree $l$, and $\delta = \sin(\omega + f) \sin i$ is the geocentric latitude. The terms $\omega, f, i$ are the satellite's argument of periapsis, true anomaly, and inclination. The third-body effect $\mathcal{R}_{3b}$ is modeled up to the fourth order via Legendre polynomials in terms of parallactic ratio $\xi = a/r_{3b}$ [10, 18]:

$$\mathcal{R}_{3b} = \frac{\mu_{3b}}{r_{3b}} \sum_{l=2}^{4} \xi^l \left(\frac{r}{a}\right)^l P_l[\cos S], \tag{3}$$

Where $\mu_{3b}$ is the third-body's gravitational parameter, $r_{3b}$ is the magnitude of the third-body vector with respect to the central planet, $a$ is the satellite semi-major axis, and $\cos S = \hat{\mathbf{r}} \cdot \hat{\mathbf{r}}_{3b}$, with $S$ the angle between the satellite and the third-body [10, 14]. Now, the spacecraft position vector is expressed in the perifocal frame: $\hat{\mathbf{r}} = \hat{\mathbf{P}} \cos f + \hat{\mathbf{Q}} \sin f$, where $\hat{\mathbf{P}} = R_3(\Omega) R_1(i) R_3(\omega) \hat{I}$ and $\hat{\mathbf{Q}} = R_3(\Omega) R_1(i) R_3\left(\omega + \frac{\pi}{2}\right) \hat{I}$. The rotation matrices $R_1(\alpha)$ and $R_3(\alpha)$ are defined as:

$$R_1(\alpha) = \begin{bmatrix} 1 & 0 & 0 \\ 0 & \cos \alpha & -\sin \alpha \\ 0 & \sin \alpha & \cos \alpha \end{bmatrix} \quad R_3(\alpha) = \begin{bmatrix} \cos \alpha & -\sin \alpha & 0 \\ \sin \alpha & \cos \alpha & 0 \\ 0 & 0 & 1 \end{bmatrix}. \tag{4}$$

Considering the third-body unit vector as $\hat{\mathbf{r}}_{3b}$, we define $\cos S$ as a direct relation of the true anomaly of the satellite: $\cos S = A_{3b} \cos f + B_{3b} \sin f$, where $A_{3b} = \hat{\mathbf{P}} \cdot \hat{\mathbf{r}}_{3b}$ and $B_{3b} = \hat{\mathbf{Q}} \cdot \hat{\mathbf{r}}_{3b}$. Under these premises, the $\mathcal{R}_{3b}$ becomes:

$$\mathcal{R}_{3b} = \frac{\mu_{3b}}{r_{3b}} \sum_{l=2}^{4} \xi^l F_l(A_{3b}, B_{3b}, r, f), \tag{5}$$

Where the second-, third-, and fourth-order terms from the polynomial expansion in $\mathcal{R}_{3b}$ have the following expressions:

$$F_2 = \left(\frac{r}{a}\right)^2 P_2[\cos S] = \frac{1}{2}\left(\frac{r}{a}\right)^2 \left(3 \cos^2 S - 1\right), \tag{6}$$

$$F_3 = \left(\frac{r}{a}\right)^3 P_3[\cos S] = \frac{1}{2}\left(\frac{r}{a}\right)^3 \left(5\cos^3 S - 3\cos S\right), \tag{7}$$

$$F_4 = \left(\frac{r}{a}\right)^4 P_4[\cos S] = \frac{1}{8}\left(\frac{r}{a}\right)^4 \left(35\cos^4 S - 30\cos^2 S + 3\right). \tag{8}$$

In this work, terms up to the fourth order have been included in $\mathcal{R}_{3b}$, resulting in: $\mathcal{R}_{3b} = \frac{\mu_{3b}}{r_{3b}}\xi^l\left(F_2 + F_3 + F_4\right)$.

**B. Averaging procedure**

To study the long-term dynamic of satellites, the short-term effects due to high-frequency variation along one orbit could be canceled out with a double averaging procedure. The first averaging (see Eq. 9) is done over the satellite's orbital period, and to do so, the disturbing function is written in terms of the eccentric anomaly and the other orbital elements of the space vehicle, following the approach in [3, 10, 18, 21]. Finally, the second averaging (see Eq. 10) is performed over the third-body orbital period by re-conducing the terms to the true anomaly, yielding a more straightforward computation.

$$\bar{\mathcal{R}} = \frac{1}{2\pi}\int_0^{2\pi} \mathcal{R}\, dM = \frac{1}{2\pi}\int_0^{2\pi} \mathcal{R}\left(1 - e\cos E\right)dE. \tag{9}$$

$$\bar{\bar{\mathcal{R}}}_{3b} = \frac{1}{2\pi}\int_0^{2\pi} \bar{\mathcal{R}}\, dM_{3b} = \frac{1}{2\pi}\int_0^{2\pi} \bar{\mathcal{R}}\, \frac{(1-e_{3b})^{3/2}}{(e_{3b}\cos f_{3b} + 1)^2}\, df_{3b}. \tag{10}$$

In the following, the subscripts $\bullet_\odot$ and $\bullet_\mathbb{C}$ are introduced for quantities related to the Sun and Moon, respectively. A complete step-by-step procedure description can be found in [22].

*1. Single-averaged disturbing function*

The single-averaged disturbing functions have been derived analytically and compared with literature results [3, 10, 14, 21]. The *single-averaged $J_2$*, in Eq. (11), depends only on the semi-major axis $a$, eccentricity $e$, and orbit inclination $i$. The *single-averaged third-body disturbing function*, in Eq. (12) also depends on the terms $A_{3b}$ and $B_{3b}$.

$$\bar{\mathcal{R}}_{J_2} = \frac{1}{2\pi}\int_0^{2\pi} \mathcal{R}_{J_2}\, dM = \frac{\mu J_2 R_\alpha^2}{8\, a^3(1-e^2)^{3/2}}(1 + 3\cos 2i). \tag{11}$$

$$\begin{aligned}\bar{\mathcal{R}}_{3b} = &-\frac{\mu_{3b}}{4r_{3b}}\xi^2\left(3A_{3b}^2\left(4e^2+1\right) - 3B_{3b}^2\left(e^2-1\right) - 3e^2 - 2\right) - \frac{5\mu_{3b}}{16r_{3b}}\xi^3 A_{3b}e\Big(-5A_{3b}^2\left(4e^2+3\right) \\ &+ 15B_{3b}^2\left(e^2-1\right) + 9e^2 + 12\Big) - \frac{3\mu_{3b}}{64r_{3b}}\xi^4\Big(35A_{3b}^4\left(8e^4+12e^2+1\right) - 10A_{3b}^2\left(7B_{3b}^2\left(6e^4-5e^2-1\right)\right.\\ &\left.+ 18e^4 + 41e^2 + 4\right) + 35B_{3b}^4\left(e^2-1\right)^2 + 10B_{3b}^2\left(3e^4+e^2-4\right) + 15e^4 + 40e^2 + 8\Big).\end{aligned} \tag{12}$$

The final expression for the single-averaged Hamiltonian representation depends on the zonal and third-body terms:

$$\bar{\mathcal{H}} = -\frac{\mu}{2\,a} - \bar{\mathcal{R}}_{J_2} - \bar{\mathcal{R}}_\odot - \bar{\mathcal{R}}_\mathbb{C}. \tag{13}$$

*2. Double-averaged disturbing function*

The double-averaging is now applied to the third-body disturbing function, which still depends on the true anomaly $f_{3b}$, via the parameters $A_{3b}$ and $B_{3b}$, and the position $r_{3b}$. First, the *Sun double-averaged function* is recovered by expressing the Sun position vector in the equatorial frame of the central body using the ecliptic longitude $l$ and the planet's obliquity of the ecliptic $\epsilon$. The coefficients $A_\odot$ and $B_\odot$ are [14]:

$$A_\odot = \cos l \,(\cos \omega \cos \Omega - \sin \omega \sin \Omega \cos i) + \sin l \,(\cos \epsilon \,(\cos \omega \sin \Omega + \cos \Omega \sin \omega \cos i) + \sin \epsilon \sin \omega \sin i)$$
$$B_\odot = -\cos l \,(\cos \Omega \sin \omega + \cos \omega \sin \Omega \cos i) + \sin l \,(\cos \epsilon \,(\sin \omega \sin \Omega + \cos \omega \cos \Omega \cos i) + \sin \epsilon \cos \omega \sin i) \quad (14)$$

For most planets, the eccentricity of the orbit around the Sun can be approximated to the circular case as a first preliminary approximation. In addition, since $l$ of the Sun varies in one year, during the motion of each planet along its orbit, from 0° to 360°, as a first approximation, the double averaging is done over $l$ instead of $f$. The second- and fourth-order double-averaged expressions are reported in Eqs. (15) and (16), similar to [3, 18, 21]. Since the circular case is adopted [18], the third-order term is null: $\bar{\bar{\mathcal{R}}}_{\odot,3} = 0$.

$$\bar{\bar{\mathcal{R}}}_{\odot,2} = \frac{\mu_\odot a^2}{16 r_\odot^3}\Big[-8 - 12 e^2 + 3(2 + 3e^2 + 5e^2 \cos 2\omega) \cos^2 \Omega - 15 e^2 \sin 2\omega \sin 2\Omega \cos i + $$
$$- 3(-2 - 3e^2 + 5e^2 \cos 2\omega) \sin^2 \Omega \cos^2 i - \Big(-3(\cos \epsilon \sin \omega \sin \Omega + \cos \epsilon \cos \omega \cos \Omega \cos i +$$
$$+ \cos \omega \sin \epsilon \sin i)^2 + 3 e^2(-\cos \epsilon \sin \omega \sin \Omega + \cos \epsilon \cos \omega \cos \Omega \cos i + \cos \epsilon \cos \omega \sin i) +$$
$$- (3 + 12 e^2)(\cos \epsilon \cos \omega \sin \Omega + \cos \epsilon \cos \Omega \sin \omega \cos i + \sin \epsilon \sin \omega \sin i)^2 \Big)\Big]. \quad (15)$$

$$\bar{\bar{\mathcal{R}}}_{\odot,4} = \frac{3 a^4 \mu_\odot}{131072 a_\odot^5}\Big(215040\, e^4 \cos^4 \omega \cos^4 \Omega + 26880\, e^4 \cos^4 \Omega \sin^4 \omega + 215040\, e^4 \cos^4 \omega \cos^4 \epsilon_\oplus \sin^4 \Omega$$
$$+ 26880\, e^4 \cos^4 \epsilon_\oplus \sin^4 \omega \sin^4 \Omega - 322560\, e^4 \cos^2 \omega \cos^4 \epsilon_\oplus \sin^2 \omega \sin^4 \Omega + ...\Big) \quad (16)$$

The full expression for the fourth-order term is lengthy and not reported in this paper for conciseness. The complete expression of $\bar{\bar{\mathcal{R}}}_{\odot,4}$ is provide in Appendix D.2 of reference [22]. For the particular case of the Earth's Moon, non-null eccentricity has been considered, yielding a different derivation. Unlike most previous works, the *Moon double-averaged function* is computed in the equatorial frame and not in the Moon plane, as in [13, 18]. The terms $A_\mathbb{C}$ and $B_\mathbb{C}$ are:

$$A_\mathbb{C} = \sin \omega \Big((\cos i \cos i_\mathbb{C} \, \cos(\Omega - \Omega_\mathbb{C}) + \sin i \sin i_\mathbb{C}) \sin(\omega_\mathbb{C} + f_\mathbb{C}) - \cos i \cos(\omega_\mathbb{C} + f_\mathbb{C}) \sin(\Omega - \Omega_\mathbb{C})\Big) +$$
$$+ \cos \omega \Big(\cos(\omega_\mathbb{C} + f_\mathbb{C}) \cos(\Omega - \Omega_\mathbb{C}) + \cos i_\mathbb{C} \, \sin(\omega_\mathbb{C} + f_\mathbb{C}) \sin(\Omega - \Omega_\mathbb{C})\Big),$$
$$B_\mathbb{C} = \cos \omega \Big((\cos i \cos i_\mathbb{C} \, \sin(\Omega - \Omega_\mathbb{C}) + \sin i \sin i_\mathbb{C}) \sin(\omega_\mathbb{C} + f_\mathbb{C}) - \cos i \cos(\omega_\mathbb{C} + f_\mathbb{C}) \sin(\Omega - \Omega_\mathbb{C})\Big) +$$
$$- \sin \omega \Big(\cos(\omega_\mathbb{C} + f_\mathbb{C}) \cos(\Omega - \Omega_\mathbb{C}) + \cos i_\mathbb{C} \, \sin(\omega_\mathbb{C} + f_\mathbb{C}) \sin(\Omega - \Omega_\mathbb{C})\Big). \quad (17)$$

The derivation is performed in terms of the true anomaly: $dM_{\mathbb{C}} = (1-e_{\mathbb{C}})^{3/2}/(e_{\mathbb{C}} \cos f_{\mathbb{C}} + 1)^2$, yielding to the double-averaged expressions in Eqs. (18), (19), and (20). The quantity $\Delta\Omega$ is the difference between the spacecraft and Moon's right ascension: $= \Omega - \Omega_{\mathbb{C}}$. Differently from [14, 18], the eccentricity of the Moon's orbit is retained in this work. Consequently, the potential odd terms are not null in this case.

$$\bar{\bar{\mathcal{R}}}_{\mathbb{C},2} = \frac{\mu_{\mathbb{C}}}{32 a_{\mathbb{C}}^3 (1-e_{\mathbb{C}}^2)^{3/2}} \Big( \big( (6+9e^2)\cos 2\Delta\Omega + 6(2+3e^2+5e^2\cos 2\omega) \big) \sin^2 \Delta\Omega \cos^2 i_{\mathbb{C}} +$$
$$+ 6(2+3e^2-5e^2\cos 2\omega)\big( (\cos\Delta\Omega \cos i_{\mathbb{C}} \cos i + \sin i_{\mathbb{C}} \sin i)^2 + \sin^2\Delta\Omega \cos^2 i \big) +$$
$$+ 5\big(-2-3e^2+6e^2\big(\cos^2\Delta\Omega\cos 2\omega + (\sin\Delta\Omega\sin 2i_{\mathbb{C}}\sin i - \sin 2\Delta\Omega \sin^2 i_{\mathbb{C}} \cos i)\big)\sin 2\omega\big) \Big),$$
(18)

$$\bar{\bar{\mathcal{R}}}_{\mathbb{C},3} = \frac{\mu_{\mathbb{C}} 15 a^3 e\, e_{\mathbb{C}}}{512 a_{\mathbb{C}}^4 (1-e_{\mathbb{C}}^2)^{5/2}} \Big( \cos\omega \cos\omega_{\mathbb{C}} \cos\Delta\Omega (35 e^2 \cos(2(\Delta\Omega+\omega)) + 35 e^2 \cos(2(\omega-\Delta\Omega))+$$
$$+ 10(e^2+6)\cos(2\Delta\Omega) + 70 e^2 \cos 2\omega - 86 e^2 - 68) - 20\sin^2 i \cos\omega (7e^2\cos 2\omega - 5e^2 - 2)\sin^2 i_{\mathbb{C}}$$
$$\cos\omega_{\mathbb{C}} \cos\Delta\Omega + \sin i \sin\omega \sin i_{\mathbb{C}} \sin\omega_{\mathbb{C}} (35 e^2 \cos(2(\Delta\Omega+\omega)) + 35 e^2 \cos(2(\omega-\Delta\Omega))+$$
$$+ 10(5e^2+2)\cos(2\Delta\Omega) + 70 e^2 \cos 2\omega - 46 e^2 - 108) + ... \Big),$$
(19)

$$\bar{\bar{\mathcal{R}}}_{\mathbb{C},4} = \frac{3 a^4 (15 e^4 + 40 e^2 + 8)(3 e_{\mathbb{C}}^2 + 2) \mu_{\mathbb{C}}}{128 a_{\mathbb{C}}^5 (1-e_{\mathbb{C}}^2)^{7/2}} + \frac{9 a^4 (e_{\mathbb{C}}^2+2)\mu_{\mathbb{C}}}{1024 a_{\mathbb{C}}^5 (1-e_{\mathbb{C}}^2)^{7/2}} \Big( 35(e^2-1)^2 (\cos i \cos\omega$$
$$(\cos i_{\mathbb{C}} \cos\omega_{\mathbb{C}} \cos\Delta\Omega + \sin\omega_{\mathbb{C}} \sin\Delta\Omega) + \sin i \cos\omega \sin i_{\mathbb{C}} \cos\omega_{\mathbb{C}} - \sin\omega \cos i_{\mathbb{C}} \cos\omega_{\mathbb{C}} \sin\Delta\Omega$$
$$+ \sin\omega \sin\omega_{\mathbb{C}} \cos\Delta\Omega)^4 + 35(8e^4 + 12 e^2 + 1)(\cos\omega \cos i_{\mathbb{C}} \cos\omega_{\mathbb{C}} \sin\Delta\Omega+$$
$$+ \cos i \sin\omega(\cos i_{\mathbb{C}} \cos\omega_{\mathbb{C}} \cos\Delta\Omega + \sin\omega_{\mathbb{C}} \sin\Delta\Omega) + ... \Big).$$
(20)

The full expressions for the third- and fourth-order terms are lengthy and not reported in this paper for conciseness. The complete expression of $\bar{\bar{\mathcal{R}}}_{\odot,4}$ is provided in Appendix C.3 of reference [22]. The final expressions of the double-averaged potential for the Sun and Moon are reported in Eq. (21). Considering the satellite semi-major axis $a$ as a constant of motion, the double-average potential is a function of the spacecraft's orbital elements and the physical properties of the third body. For the Moon case, the potential depends on the node $\Omega_{\mathbb{C}}$ and inclination $i_{\mathbb{C}}$, which are time-varying quantities in the equatorial representation.

$$\bar{\bar{\mathcal{R}}}_{\odot} = \bar{\bar{\mathcal{R}}}_{\odot}(a, e, i, \omega, \Omega, -; \epsilon, r_{\odot}, \mu_{\odot})$$
$$\bar{\bar{\mathcal{R}}}_{\mathbb{C}} = \bar{\bar{\mathcal{R}}}_{\mathbb{C}}(a, e, i, \omega, \Omega, -, a_{\mathbb{C}}, e_{\mathbb{C}}, i_{\mathbb{C}}, \omega_{\mathbb{C}}, \Omega_{\mathbb{C}}, -; \mu_{\mathbb{C}}).$$
(21)

The final expression for the double-averaged Hamiltonian representation depends on the zonal and third-body terms:

$$\bar{\bar{\mathcal{H}}} = -\frac{\mu}{2a} - \bar{\bar{\mathcal{R}}}_{J_2} - \bar{\bar{\mathcal{R}}}_{\odot} - \bar{\bar{\mathcal{R}}}_{\mathbb{C}}.$$
(22)

## 3. Single- and double-averaged models validation

After the derivation of single- and double-averaged potentials, the model's accuracy is compared to the non-averaged expression of the dynamics (see Eq. (22)). The reference Earth's orbit under study is an INTEGRAL-like orbit with initial conditions on 22/03/2013 [14]. The parameters for setting up the simulation are reported in Table 1, including the orbital elements in Earth's J2000 frame. The dynamic is integrated via an ordinary differential equation solver for stiff problems, with a variable step and order (1 to 5). Fig. 1 shows the comparison of the time evolution of the Keplerian elements and the altitude of periapsis $h_p$ between full, single- and double-averaged models for about 3.5 years. One can note that the averaging procedure correctly approximates the long-term dynamic. After the double-averaging procedure, the Hamiltonian is a non-autonomous 2-DOF system, with the time dependencies stemming from the third bodies' ephemeris. An additional reduction of the Hamiltonian description to 1-DOF is required to get a two-dimensional phase space for a fully-analytical design of disposal maneuvers (see Section III).

**Table 1  Initial conditions for simulating an INTEGRAL-like orbit.**

| Parameters | Unit | Value |
|---|---|---|
| Orbit's Keplerian el. $(a, e, i, \Omega, \omega, M)$ | (km, -, rad, rad, rad, rad) | $[87705.22, 0.8766, 1.0739, 2.2516, 4.6385, 4.15]$ |
| Earth's gravitational constant | km$^3$/s$^2$ | $3.9860 \cdot 10^5$ |
| Earth's mean equatorial radius | km | $6.3782 \cdot 10^3$ |
| Earth's oblateness $J_2$ | - | 0.001082628 |
| Obliquity of the ecliptic $\epsilon$ | rad | 0.4091 |
| Sun's gravitational constant $\mu_\odot$ | km$^3$/s$^2$ | $1.3271 \cdot 10^{11}$ |
| Sun-Earth distance | km | $1.4962 \cdot 10^8$ |
| Moon's gravitational constant $\mu_\mathbb{C}$ | km$^3$/s$^2$ | $4.903 \cdot 10^3$ |
| Moon's Keplerian el.* $(a, e, i, \Omega, \omega)$ | (km,-,rad,rad,rad) | $[3.8440 \cdot 10^5, 0.4877, 0.0549, 4.4685, 0.1013]$ |

*Analytical ephemeris of the Moon from [20]

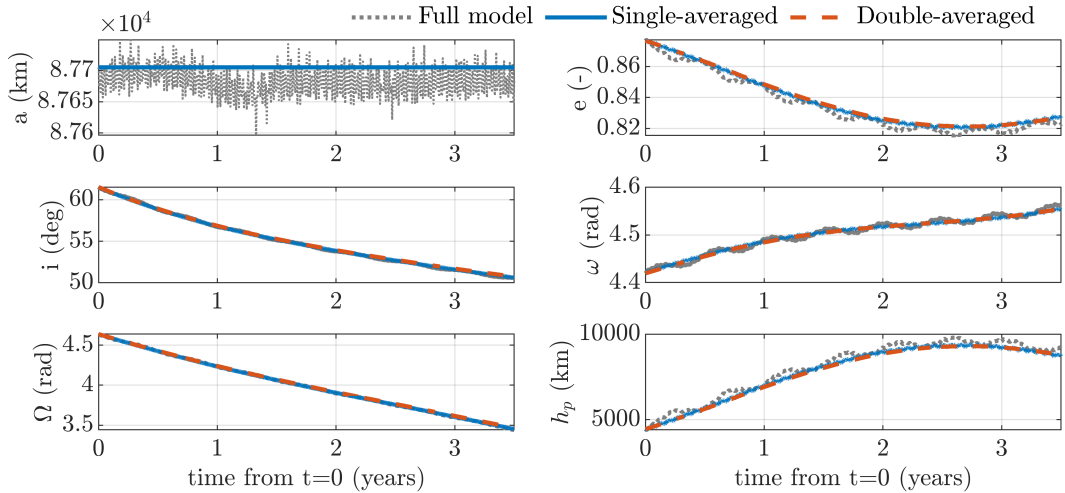

**Fig. 1  Comparison between the full (exact), single-, and double-averaged models.**

## C. Reduced Hamiltonian Formulation

The approach to reducing the Hamiltonian to a 1-DOF is based on an averaging procedure over the satellite node $\Omega$, called the *elimination of the node*. Only the second-order contributions are fully reported in this article, while the complete derivation has been described in [22].

### 1. Elimination of the node

The elimination of the node consists of the averaging on $\Omega$ over one orbital period. The following expression for the reduced Hamiltonian $\hat{\mathcal{H}}$ has been computed for the second-order term:

$$\hat{\mathcal{H}} = \frac{1}{2\pi} \int_0^{2\pi} \bar{\bar{\mathcal{H}}} \, d\Omega = -\frac{\mu}{2a} - \frac{J_2 \mu R_\oplus^2 (3\cos 2i + 1)}{8a^3 (1-e^2)^{3/2}} - \frac{\mu_\mathbb{C} \, a^2}{32 \, a_\mathbb{C}^3 \, (1-e_\mathbb{C}^2)^{3/2}} \Big( 15 \, e^2 \cos \omega (2 \sin^2 i \sin^2 i_\mathbb{C} +$$
$$+ (-1 + \cos^2 i)(1 + \cos^2 i_\mathbb{C})) - (2 + 3e^2)(-5 + +3 \cos^2 i (1 + \cos^2 i_\mathbb{C}) + 6 \sin^2 i \sin^2 i_\mathbb{C}) \Big) +$$
$$- \frac{\mu_\odot a^2}{64 \, r_\odot^3} \Big( 10(-2 - 3e^2 + 3e^2 \cos \omega) - 3(3 + \cos 2\epsilon) \cos^2 i (-2 - 3e^2 + 5e^2 \cos 2\omega) 6 \cos^2 \epsilon$$
$$(2 + 3e^2 + 3e^2 \cos 2\omega) - 12(-2 - 3e^2 + 5e^2 \cos 2\omega) \sin^2 \epsilon \sin^2 i \Big).$$
(23)

The third- and fourth-order terms have not been included in the Hamiltonian expression for conciseness of the representation, but they have been considered during the simulations. Considering constant orbital elements of the perturbing bodies, the reduced Hamiltonian is a function of $(a, e, i, \omega)$:

$$\hat{\mathcal{H}} = \hat{\mathcal{H}}(a, e, i, \omega, -, -; a_\mathbb{C}, e_\mathbb{C}, i_\mathbb{C}, \omega_\mathbb{C} \, \epsilon, r_\odot, J_2, \mu, \mu_\odot, \mu_\mathbb{C}, R_\oplus) \longrightarrow \hat{\mathcal{H}}(a, e, i, \omega) \quad (24)$$

Thanks to the time independency of Eq. 24, the Kozai parameter representation can be applied to the model [2]. The Kozai parameter was defined in [2] as a constant of motion, function of eccentricity and inclination of an orbit: $\Theta_{kozai} = (1 - e^2) \cos^2 i$. Substituting the inclination dependency on the eccentricity from $\Theta$ in Eq. 24, and considering a constant semi-major $a = a_0$, the reduced Hamiltonian becomes a function of $(e, \omega)$ only: $\hat{\mathcal{H}}(-, e, -, \omega)$. The reduced Hamiltonian representation $\hat{\mathcal{H}}$ provides two-dimensional phase-space maps in terms of a constant semi-major axis and the initial conditions of the satellite's orbit. The phase space maps are produced by computing the contour plot of the Hamiltonian function, defined as:

$$\mathcal{F} = \hat{\mathcal{H}}(e, \omega) - \hat{\mathcal{H}}_0(e_0, \omega_0), \quad (25)$$

Where $(e_0, \omega_0)$ are the initial conditions (only the dynamical dependencies are reported in Eq. (25)). The two-dimensional phase-space maps are used to design the fully-analytical maneuvers in Section III.B. Note that the Kozai parameter would not be constant for the Earth-Moon-Sun system. Particularly, $\Theta_{kozai,0}$ has been derived by [2] assuming a non-inclined third body. This is invalid for the Moon's orbit, and the node elimination introduces an approximation.

*2. Reduced Hamiltonian model validation*

The reduced Hamiltonian $\hat{\mathcal{H}}$ has been compared with the full, single-, and double-averaged for validation. Two situations have been analyzed. First, we consider a satellite orbiting the Earth with initial conditions of Table 1, under the $J_2$, Moon, and Sun effects. We also assume a small area-to-mass ratio ($\ll 1$) to neglect the solar radiation pressure. Fig. 2 shows a non-accurate approximation of the reduced Hamiltonian model, caused by the approximation introduced by the 1-DOF reduction (see the discussion in Section IV.B.2). Second, we introduce the assumption that the third bodies lie on the Earth's equatorial plane (i.e., no inclination of the Moon over the equator, $i_{\mathbb{C}} = 0$ deg) for initial conditions of Table 1. This situation represents a system where the third body lies on the equatorial plane of the central planet (e.g., a probe around Venus or Jupiter's moons, Europa, Ganymede, and Callisto). In this case, the 1-DOF model accurately describes the long dynamic dynamics for a non-inclined third-body, as shown in Fig. 3.

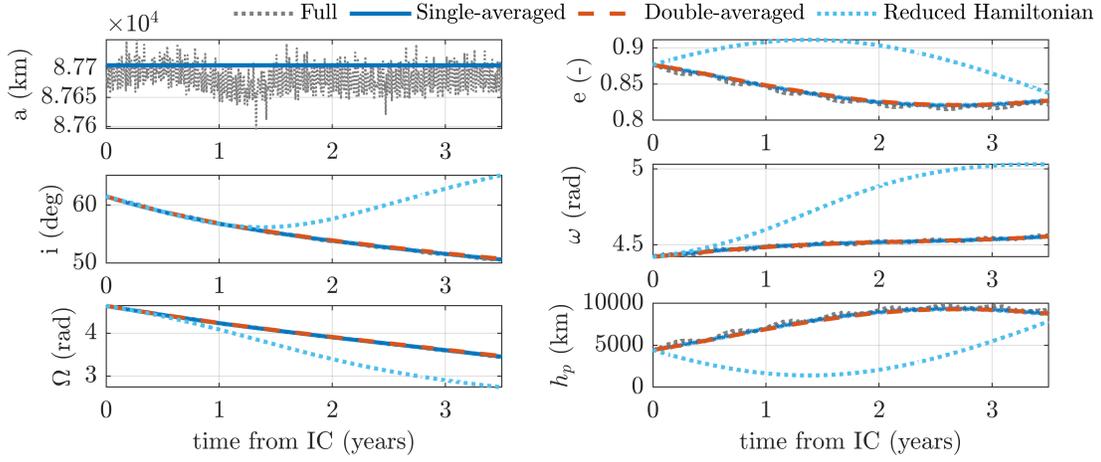

**Fig. 2  Comparison of full, single/double-averaged, and reduced Hamiltonian models (inclined Moon).**

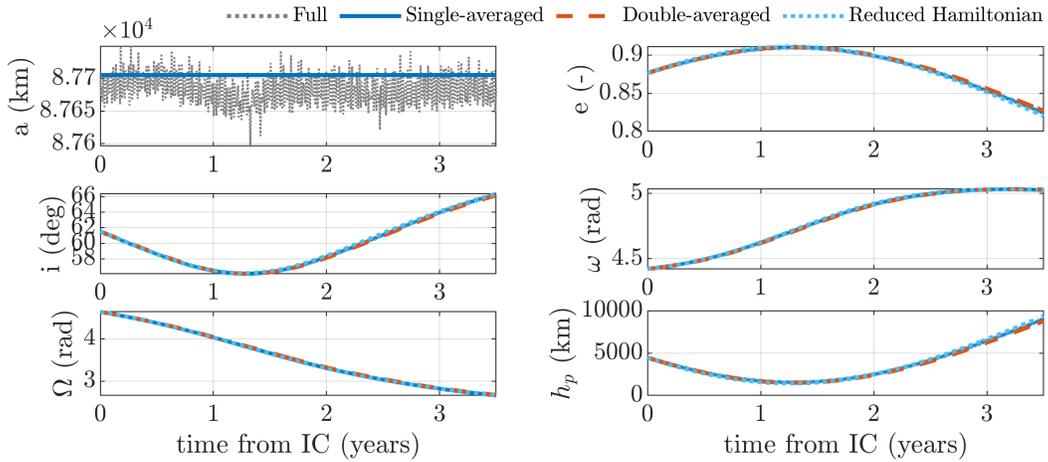

**Fig. 3  Comparison of full, single/double-averaged, and reduced Hamiltonian models (non-inclined Moon).**

After validating the double-averaged and the reduced Hamiltonian models against the full (exact) propagation, the manuscript presents an innovative strategy for end-of-life maneuvers design based on the phase-space representation of the spacecraft dynamics from the reduced Hamiltonian model $\hat{\mathcal{H}}$.

## III. Design of Optimal Disposal Maneuvers in the Phase-Space

This section presents the design of optimal disposal maneuvers in the atmosphere, exploiting the phase-space representation. The procedure consists of an impulsive $\Delta V$ maneuver to target specific orbital elements to reach the disposal condition. Differently from the semi-analytical model in [23], this work proposes a novel fully-analytical approach to design the disposal maneuver based on the reduced Hamiltonian formulation. The disposal in the atmosphere requires the altitude of the periapsis below a specific threshold. Starting from the relation among the semi-major axis, altitude of the periapsis, and eccentricity of an orbit: $e = 1 - \frac{h_p + R_\alpha}{a}$, the eccentricity value corresponding to the altitude for atmospheric re-entry is identified, namely the critical eccentricity $e_{cr}$. The altitude threshold is typically set around 120 km for an Earth satellite [24]. The $\Delta V$ maneuver is designed in the local orbital frame, i.e., the Local Vertical/Local Horizontal (LV/LH), defined by: x-axis aligned with the velocity vector, y-axis in the direction of the orbital angular momentum, and the z-axis completes the orthogonal frame (i.e., toward the nadir). A maneuver in the LV/LH frame can be described using two angles $\alpha$ (in-plane x-z) and $\beta$ (out-of-plane y) and the magnitude $\Delta V$:

$$\Delta \mathbf{V}|_{LV/LH} = \Delta V \left[ \cos\alpha \cos\beta; \quad \sin\beta; \quad \sin\alpha \cos\beta \right]' \tag{26}$$

After the application of the impulsive $\Delta V$, the new Keplerian elements $\texttt{kep}^+$ are computed from the finite variation of initial conditions $\texttt{kep}^-$ through the Gauss planetary equations in terms of impulsive $\Delta \mathbf{V}|_{LV/LH}$ [25] (see [20] for a full description of the Gauss planetary equations ). Then, the $\texttt{kep}^+$ is propagated to check for atmospheric re-entry conditions under two different approaches: first, a semi-analytical propagation based on the double-averaged model; second, a fully-analytical propagation based on the reduced Hamiltonian model.

### A. Semi-analytical approach

The Keplerian elements $\texttt{kep}^+$ are propagated in time using the double-averaged model under a semi-analytical propagation [18]. A set of $\Delta \mathbf{V}$ with different magnitudes and orientations is applied to the initial conditions $\texttt{kep}^-$. During the dynamical propagation $\texttt{kep}^+$, the eccentricity value is compared with the critical value $e_{cr}$. Only the $\Delta Vs$ that result in the atmospheric re-entry are retained in the solution (i.e., if $e_{cr}$ is reached). For those conditions, the time evolution in terms of altitude of periapsis $h_p(t)$ is computed, and the minimum altitude $h_{p,min}$ for atmospheric re-entry is achieved at a specific time instant, similar to the work in [26]. The true anomaly $f$ of the impulsive $\Delta V$ can be optimized a-posteriori since the double-averaged model is independent of the true anomaly $f$.

## B. Fully-analytical approach

An innovative fully-analytical procedure has been proposed starting from the reduced Hamiltonian model. The corresponding phase space provides an intuitive visualization of the maneuver effect. We use the expression in Eq. 25 to obtain the two-dimensional maps: the phase space corresponds to its contour plot. An example of the phase space map is reported in Fig. 4 (a), depicting an INTEGRAL-like orbit. The value for the critical eccentricity $e_{cr}$ to reach re-entry conditions is represented in Fig. 4 (b) with a blue rectangle. The $\Delta V$ modifies the INTEGRAL trajectory in the phase space so that the critical eccentricity condition is reached (see the red trajectory). After applying a $\Delta V$, the $e_{cr}$ is achieved if the maximum eccentricity of the trajectory evolution in the phase space is equal or higher to $e_{cr}$. The maximum eccentricity value can be computed analytically from Eq. 25, as a stationary point as a function of $\omega$ (maximum of $\mathcal{F}$). This procedure is more computationally efficient than the semi-analytical one, as no numerical integration is required. Depending on the impulsive $\Delta V$, for a given magnitude and directions $\alpha$ and $\beta$, we obtain different conditions for Keplerian elements $\text{kep}^+$. The Hamiltonian contour line could translate up or down: for a reduction of $a$, the phase space translates towards higher eccentricity values, enhancing the disposal condition, and vice-versa. Fig 4 (b) shows how the Hamiltonian phase space changes after the $\Delta V$ is applied for an INTEGRAL-like trajectory in the Earth-Moon-Sun system. The black trajectory represents the evolution from the initial conditions, while the red one describes the evolution of the orbital conditions after the maneuver. Since the initial trajectory (in light blue) is not tangent to the disposal condition, i.e. $e_{cr}$, an impulsive maneuver is applied to enhance the effect of Earth's oblateness $J_2$ and third-body perturbations. However, once the $\Delta V$ is applied, the new orbital elements of the satellites result in a new trajectory (in red), tangent to the critical condition $e_{cr}$. The re-entry could not be immediate, but the trajectory propagates in time until the critical eccentricity condition is reached, leading to atmospheric re-entry.

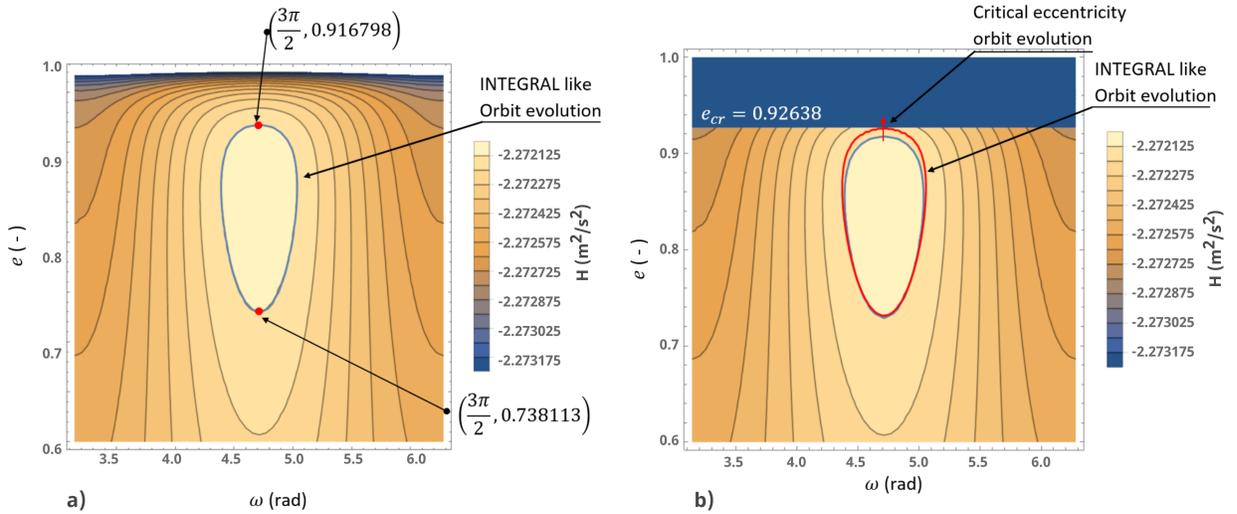

**Fig. 4** **2D phase-space in Earth-Moon-Sun system for INTEGRAL-like trajectory.**

## C. Optimisation procedure

An optimization procedure has been implemented to compute the optimal maneuver in terms of $\Delta V$ magnitude, direction, and true anomaly $f$. A similar approach of [14, 18] has been adopted, implementing some improvements as presented in this section. The optimization aims to determine the minimum impulsive $\Delta V$ to perform an atmospheric re-entry for a specific scenario. The optimization parameters are defined as $\mathbf{x} = [\alpha, \beta, \Delta V, f_m]$. A multi-objective optimization has been implemented [27], where the optimal solution targets the critical eccentricity and the minimum $\Delta V$. The former condition is related to the altitude of the periapsis for the re-entry, and it has a higher relative importance than the $\Delta V$, related to propellant consumption. The cost function for the optimal control problem is selected as:

$$C = \frac{1}{2}\left(KC_{h_p} + WC_{\Delta V}\right), \qquad (27)$$

Where $K = 1$ and $W = 1 \times 10^{-2}$ are the weighting constants for the optimization. The weighting constants are selected to grant the convergence in terms of target periapsis for the re-entry condition and minimum $\Delta V$. The first term $C_{h_p}$ minimizes the altitude of periapsis to target the re-entry altitude. It depends on the actual and critical values of $h_p$, and $h_{p,cr}$, where the latter value corresponds to the eccentricity $e_{cr}$:

$$C_{h_p} = \max\left(\frac{h_{p,min} - h_{p,cr}}{h_{p,cr}}, 0\right)^2. \qquad (28)$$

Differently from the cost function used in [26], the variation in the periapsis altitude is divided by the target altitude to introduce a weighting coefficient for the objective function, resulting in an a-dimensional cost function. The second objective of the optimization is to maintain the $\Delta V$ cost the smallest as possible, as the onboard fuel at the end of the mission is typically extremely low. The cost function for the optimal $\Delta V$ impulse is defined as:

$$C_{\Delta V} = \left(\frac{\Delta V}{\sigma_v}\right)^2, \qquad (29)$$

Where $\sigma_v$ is set equal to 1 km/s to have an a-dimensional cost function. Differently from [18, 26], we introduced in the cost function the weighting factors ($h_{p,target}, \sigma_v$). The following optimal control problem has been set up:

$$\begin{aligned}\texttt{minimize} \quad & C = \frac{1}{2}\left(KC_{h_p} + WC_{\Delta V}\right) \\ \texttt{subject to} \quad & -\pi \leq \alpha \leq \pi; \quad -\pi/2 \leq \beta \leq pi/2; \quad \Delta V_{\min} \leq \Delta V \leq \Delta V_{\max}; \quad 0 \leq f_m \leq 2\pi\end{aligned} \qquad (30)$$

Where $\Delta V_{\min}$ and $\Delta V_{\max}$ are the minimum and maximum magnitude for the maneuver. The optimization is performed with a multi-start method. It exploits local searching procedures from random initial solutions, including lower and upper boundaries [28]. It generates multiple local solutions starting from various initial points in an attempt to find the global minima inside the boundaries, and it is based on constrained nonlinear programs. Specifically, the `MultiStart` algorithm in MATLAB® has been considered. The solution provides information on the global minimum and the initial conditions that lead to the minimum. The advantage of using the `MultiStart` solver relies on the identification of global optima and a faster convergence compared to a genetic algorithm (see Ref. [22]).

## IV. Results

The semi- and fully-analytical methods have been applied to two case scenarios. As described in Section III, the former method is based on the double-averaged model, which accurately approximates the exact dynamical propagation under $J_2$ and third-body perturbations. The latter is based on the reduced Hamiltonian. As described in Section II.C, the reduced Hamiltonian accurately describes the dynamics for the case of a non-inclined third body. However, it fails to capture the dynamical evolution for the inclined case accurately. Consequently, the following test case scenarios have been considered in this work to assess the potential efficiency of the fully-analytical methods compared to the semi-analytical ones.

The first test case considers a scenario with the third body lying on the equator of the central planet. To avoid approximating the Moon effect to a non-inclined body, Venus was selected as the central planet instead of the Earth. Mission to Venus are of interest in the scientific community to improve the knowledge of its atmospheric composition and other physical properties [7, 27, 29] The Venus-Sun system has been considered including the planet's oblateness and Sun's third-body effects. The relatively small inclination of Venus' equatorial plane over the ecliptic plane (around 2.64 deg) allows the approximation of the Sun on the planet's equatorial plane [7]. The following Hamiltonian representation has been considered to derive the fully-analytical approach:

$$\mathcal{H} = -\frac{\mu}{2a} - \mathcal{R}_{J_2} - \mathcal{R}_\odot. \tag{31}$$

The validation of the model for a non-inclined third-body has been described in Fig.3.

The second test case considers an INTEGRAL-like satellite in the Earth-Moon-Sun system, under the Earth's oblateness, the Sun's and the Moon's gravitational effects. The following Hamiltonian representation was implemented:

$$\mathcal{H} = -\frac{\mu}{2a} - \mathcal{R}_{J_2} - \mathcal{R}_\odot - \mathcal{R}_\mathbb{C}. \tag{32}$$

When an inclined third body is considered, the reduced Hamiltonian method does not correctly approximate the dynamical behavior. Consequently, the analysis is performed by comparing the results of the semi- and the fully-analytical approaches. In addition, the results of both approaches are compared with the literature results for INTEGRAL satellite [14, 26].

## A. Test case 1: Venus Orbiter

This scenario describes the atmospheric re-entry of a probe on an HEO around Venus. The atmospheric interface is around 250 km [30], and the target altitude has been selected slightly lower (130 km) in the optimization problem to ensure re-entry. Moreover, the following constraints have been considered: the disposal should be provided within a 15-year window, the $\Delta v$ can vary in the interval (0 - 1.2) km/s, and $\alpha$ and $\beta$ angles in the range (0,360) deg. Table 2 reports the initial conditions for the probe around Venus and the physical properties of the central planet. The target altitude for the re-entry corresponds to a critical eccentricity condition $e_{cr} = 0.9281$. The analyses consist of two optimization simulations to compare the performances of two different maneuver points: at M1, the minimum eccentricity condition, and at M2, the maximum eccentricity condition. The maneuver was modeled with the fully-analytical and semi-analytical approaches to assess the accuracy of the results. The representation of the disposal maneuver in the phase space is shown in Fig. 5: the initial phase space is shown by the blue lines, while the final phase space is in red. It also represents the maneuver points M1 and M2. The semi- and fully-analytical optimization procedure results are reported in Table 3. Overall, the fully-analytical method is more computationally efficient than the semi-analytical one, requiring less than 10 seconds to converge to the optimal solution. Moreover, one can observe that the disposal maneuver at M2 (i.e. point at maximum eccentricity) is more expensive than the maneuver at M1 (i.e. point of minimum eccentricity) in terms of $\Delta V$. A maneuver at M2 corresponds to a direct reduction of the periapsis altitude below the disposal condition. On the other hand, the maneuver at M1 produces a decrease in the semi-major axis and inclination of the orbit, with a lower altitude of periapsis: the natural evolution under external perturbation generates a re-entry in about five years from the maneuver. Finally, Table 3 shows that the two methods provide similar results for modeling the disposal maneuver of Venus' probe, validating the relevance of the innovative fully-analytical approach for fast and accurate design.

Table 2  Initial conditions for simulating a probe around Venus.

| Parameters | Unit | Value |
|---|---|---|
| Coordinate Universal Time | - | 00:00 22-03-2013 |
| Orbit's Keplerian el. $(a, e, i, \Omega, \omega, M)$ | (km, -, rad, rad, rad, rad) | [87000.0, 0.87, 1.047, 4.42, 4.64, 2.25] |
| Venus's gravitational constant | km$^3$/s$^2$ | $3.2486 \cdot 10^5$ |
| Venus's mean equatorial radius | km | $6.0518 \cdot 10^3$ |
| Venus's oblateness $J_2$ | - | $4.458 \cdot 10^{-6}$ |
| Obliquity of the ecliptic $\epsilon$ | rad | 0.046 |
| Sun's gravitational constant $\mu_\odot$ | km$^3$/s$^2$ | $1.3271 \cdot 10^{11}$ |
| Sun's semi-major axis | km | $1.0821 \cdot 10^8$ |

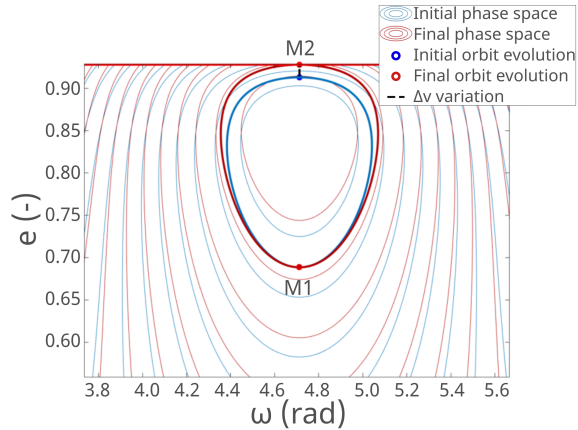

Fig. 5  Phase space trajectory in $(\omega, e)$ to target the re-entry.

Table 3   Results for $\Delta$v optimization procedure for the disposal of Venus' orbiter.

| Method | Maneuver point | $\Delta$V, m/s | Minimum altitude, km | Computational time, min |
| --- | --- | --- | --- | --- |
| Semi-analytical | M1 | 60 | 130 | ~60 |
| Semi-analytical | M2 | 84 | 130 | ~60 |
| Fully-analytical | M1 | 57 | 130 | ~3 |
| Fully-analytical | M2 | 86 | 130 | ~5 |

**B. Test case 2: INTEGRAL satellite**

The second test case considers the disposal maneuver for the INTEGRAL satellite. The target altitude is selected equal to 50 km (well below 120 km) to ensure the re-entry and minimize possible atmospheric fragmentation before the re-entry. The initial conditions for the simulation of the INTEGRAL re-entry are reported in Table 1. As for the first test case, the disposal is required within a 15-year window, and the $\Delta$v can vary in the interval $0 - 1.2$ km/s, and $\alpha$ and $\beta$ angles in the range $0 - 360$ deg. Two different analyses have been performed for this scenario. First, the cost of the disposal maneuver is evaluated at two eccentricity conditions, minimum (M1) and maximum (M2). The maneuver has been modeled with the semi- and fully analytical approaches. Figure 6 a) shows the initial and final phase space, in blue and red, respectively, and the two maneuvers at M1 and M2. Table 4 reports the results in terms of $\Delta$v, computational time, and minimum periapsis altitude. The fully analytical solution results in a minimum $\Delta$v at M2 and a maximum $\Delta$v at M1, contrary to the results of the semi-analytical one. This discrepancy is due to the lower accuracy of the reduced Hamiltonian model for the case of an inclined Moon. The second analysis consists of an optimization procedure considering a set of 20 initial conditions in terms of initial time, starting from 2013 for 25 years. We also compare the results with the outcomes of [26]. Figure 6 b) shows the optimal $\Delta$v for different initial disposal times. The blue line represents the fully-analytical solution, the green one the semi-analytical, and the red line represents the solution obtained in *Colombo et al. (2014)* [26] for the same initial conditions. As expected, the computations performed with the semi-analytical method are much more accurate than the fully-analytical one. The latter results in cheaper maneuvers when it should be more expensive and vice-versa. The behavior of the semi-analytical method is comparable with the results in *Colombo et al. (2014)* [26], even if a smaller number of initial conditions have been analyzed. The numerical results are reported in Table 5, where the best solutions in terms of $\Delta v$ and periapsis altitude are highlighted in blue. The three best options can be identified in 2014, 2023, and 2032. the cheapest option is in 2023, with a $\Delta v$ of about 47 m/s.

Table 4   Results for the disposal of INTEGRAL satellite at initial time 22-03-2013.

| Method | Maneuver point | $\Delta$V, m/s | Minimum altitude, km | Computational time, min |
| --- | --- | --- | --- | --- |
| Semi-analytical | M1 | 45.7 | 50 | ~60 |
| Semi-analytical | M2 | 97 | 50 | ~60 |
| Fully-analytical | M1 | 106 | 50 | ~3 |
| Fully-analytical | M2 | 67 | 50 | ~5 |

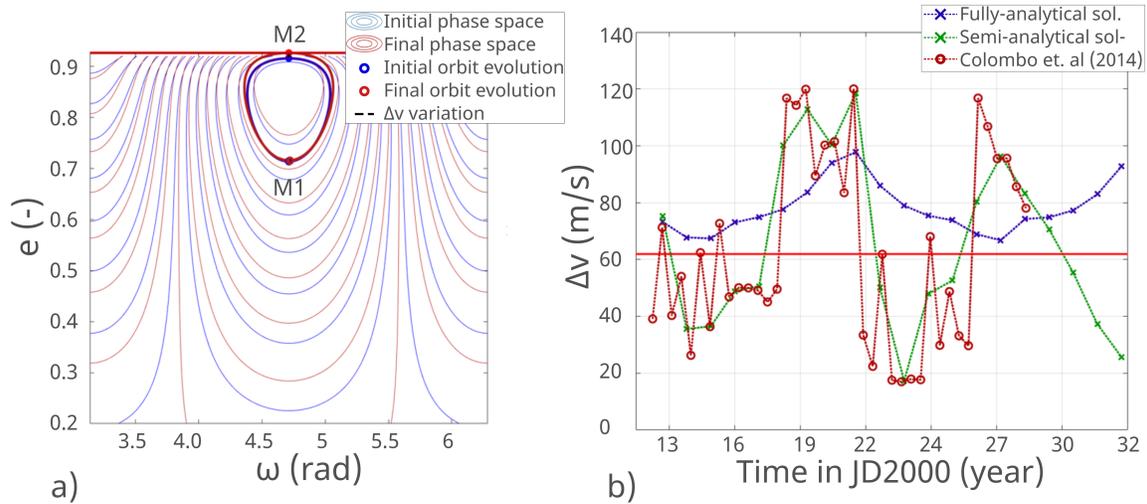

**Fig. 6  INTEGRAL satellite: a) Phase space, b) Optimization procedure.**

*1. Semi-analytical and Fully-analytical methods performances*

The comparison between the computational time for disposal options demonstrates that using a semi-analytical propagation for maneuver optimization is more expensive than a fully-analytical method based on the solution of the reduced Hamiltonian. Furthermore, the semi-analytical method could take several hours to produce optimal results, as reported in Table 6. For this reason, the optimal solution is typically computed on the ground, and then the instructions are sent to the onboard system. On the contrary, the approach based on the reduced Hamiltonian significantly reduces the computational time of the optimization procedure. The power of the fully-analytical approach relies on the computational time to solve for stationary point conditions. As shown in Table 6, it requires less than 10 min to converge to the optimal solution and about 0.02 seconds to propagate the trajectory's initial conditions for 25 years, compared to the 60 minutes and 3 seconds, respectively, for the semi-analytical one. The performances have been evaluated with a processor of 2.60 GHz and 16.0GB of RAM. To conclude, the computational time is reduced significantly, yielding the need to develop a more accurate analytical model.

Table 5  Results for the disposal of INTEGRAL satellite at initial time 22-03-2013.

| Maneuver date | Fully-analytical | | Semi-analytical | |
| --- | --- | --- | --- | --- |
| | Δv, m/s | Minimum altitude, km | Δv, m/s | Minimum altitude, km |
| 01/06/2013 | 73.3 | 34.69 | 75.2 | 50.02 |
| 04/06/2014 | 67.7 | 33.18 | 35.5 | 49.5 |
| 08/06/2015 | 67.4 | 43.45 | 36.5 | 50.3 |
| 11/06/2016 | 73.1 | 43.17 | 48.8 | 49.8 |
| 14/06/2017 | 74.9 | 53.23 | 50.6 | 49.7 |
| 18/06/2018 | 77.7 | 38.76 | 100.1 | 49.1 |
| 22/06/2019 | 83.7 | 44.37 | 112.8 | 50.2 |
| 24/06/2020 | 94.0 | 32.37 | 100.3 | 49.4 |
| 28/06/2021 | 97.8 | 35.37 | 118.5 | 49.9 |
| 02/07/2022 | 85.9 | 44.38 | 50.1 | 50.0 |
| 05/07/2023 | 78.9 | 37.43 | 17.2 | 47.8 |
| 08/07/2024 | 75.5 | 37.21 | 47.9 | 45.6 |
| 12/07/2025 | 73.8 | 37.63 | 52.6 | 48.5 |
| 15/07/2026 | 68.8 | 33.12 | 80.4 | 49.7 |
| 19/07/2027 | 66.8 | 52.19 | 96.3 | 50.8 |
| 22/07/2028 | 74.2 | 40.47 | 83.2 | 47.3 |
| 25/07/2029 | 74.8 | 43.05 | 70.6 | 50.3 |
| 29/07/2030 | 77.3 | 45.10 | 55.3 | 46.2 |
| 02/08/2031 | 83.0 | 40.85 | 37.2 | 48.8 |
| 05/08/2032 | 92.2 | 42.49 | 25.6 | 49.3 |

Table 6  Difference in computational time between the semi- and fully-analytical approaches.

| Method | Propagation time for 25 years, sec | Optimization time, min |
| --- | --- | --- |
| Semi-analytical | 3.63 | > 60 |
| Fully-analytical | 0.022 | < 10 |

*2. Problem of Node elimination for Earth-Moon-Sun system*

The results, given by the Earth-Moon-Sun model, highlight the limitations of the reduced Hamiltonian model for the inclined third-body case. Even if the reduced Hamiltonian methodology is very promising as it allows the design of optimal disposal maneuvers with no integration of the dynamics but simply by solving the 2D Hamiltonian equation, the model produces reliable results only for a system where the relative inclination of the third body upon the equator is negligible, as the Venus case. This suggests that different approaches should be investigated for the Hamiltonian reduction in the case of an inclined third body. For these cases, eliminating the satellite's node is a non-trivial process. Therefore, the complexity of such an aspect should be tackled directly in the reduction procedure of the Hamiltonian. Specifically, for the Earth-Moon-Sun system, the Moon node has a non-linear variation on the equatorial plane. The coupling effect with the satellite node causes complex secular dynamical behavior. Accordingly, the reduction procedure drops important contributions in the secular and long-term satellite evolution, reducing the model's accuracy.

## V. Conclusions

This manuscript presents a design procedure for disposal maneuvers of a probe orbiting a planet under the oblateness and third-body perturbations. As a result of the 2-D phase-space maps, this work provide a preliminary method to design disposal maneuvers for spacecraft in highly elliptical orbits. The benefit of the presented fully-analytical approach is the reduction of the computational time for delta-v maneuver optimization. For the case of non-inclined third body, the fully-analytical approach accurately describes the long-term dynamical evolution and provides comparable results to the semi-analytical procedure, already proposed in past works. Therefore, the fully-analytical approach is a promising method for systems with non-inclined third bodies. As an example, this work proposes an innovative, fast, and reliable way to design maneuvers for probes orbiting Venus or Jupiter's moons. However, the methodology proposed in this work has some limitations. The procedure followed to obtain the reduced Hamiltonian model causes a loss in accuracy when the third body is on an inclined orbit around the central planet. This is the main drawback for scenarios in the Earth-Moon-Sun system. The node elimination procedure removes complex dynamics of the Moon's third-body perturbation by eliminating the dependency on the Moon node. For this reason, the fully-analytical model is not accurate for the Earth's system and, therefore, can only be used for a very preliminary analysis of the order of magnitude of the maneuver effort.

## Funding Sources

This project has received funding from the European Research Council (ERC) under the European Union'sHorizon 2020 research and innovation program (grant agreement No 679086 - COMPASS).